\documentclass[a4paper,fleqn,usenatbib]{mnras}

\usepackage{newtxtext,newtxmath}
\usepackage[T1]{fontenc}
\usepackage{ae,aecompl}

\usepackage{graphicx}
\usepackage{url,hyperref}
\usepackage{txfonts}
\usepackage{footmisc}

\title[H.E.S.S. observations of PKS 1830-211]{H.E.S.S. observations of the flaring gravitationally lensed galaxy PKS 1830-211}

\author[H.E.S.S. Collaboration]
{\Large\parbox{\textwidth}{
H.E.S.S. Collaboration,
H.~Abdalla$^{1}$,
F.~Aharonian$^{3,4,5}$,
F.~Ait~Benkhali$^{3}$,
E.O.~Ang\"uner$^{19}$,
M.~Arakawa$^{39}$,
C.~Arcaro$^{1}$,
C.~Armand$^{23}$,
M.~Arrieta$^{14}$,
M.~Backes$^{8,1}$,
M.~Barnard$^{1}$
% ...
}
}

\makeatletter
\renewcommand*{\@fnsymbol}[1]{\ifcase#1\@arabic{#1}\fi}
\makeatother 

\date{Accepted XXX. Received YYY; in original form ZZZ}

\pubyear{2019}

\begin{document}
\label{firstpage}
\pagerange{\pageref{firstpage}--\pageref{lastpage}}
\maketitle

%\input{authors.tex}
%$^{*}$Corresponding authors
\clearpage

\parbox{\textwidth}
{
$^{1}$Centre for Space Research, North-West University, Potchefstroom 2520, South Africa\\ 
$^{2}$Universit\"at Hamburg, Institut f\"ur Experimentalphysik, Luruper Chaussee 149, D 22761 Hamburg, Germany\\
$^{3}$Max-Planck-Institut f\"ur Kernphysik, P.O. Box 103980, D 69029 Heidelberg, Germany\\ 
$^{4}$Dublin Institute for Advanced Studies, 31 Fitzwilliam Place, Dublin 2, Ireland\\ 
$^{5}$National Academy of Sciences of the Republic of Armenia,  Marshall Baghramian Avenue, 24, 0019 Yerevan, Republic of Armenia \\ 
$^{6}$Yerevan Physics Institute, 2 Alikhanian Brothers St., 375036 Yerevan, Armenia\\
$^{7}$Institut f\"ur Physik, Humboldt-Universit\"at zu Berlin, Newtonstr. 15, D 12489 Berlin, Germany\\ 
$^{8}$University of Namibia, Department of Physics, Private Bag 13301, Windhoek, Namibia\\ 
$^{9}$GRAPPA, Anton Pannekoek Institute for Astronomy, University of Amsterdam,  Science Park 904, 1098 XH Amsterdam, The Netherlands\\ 
$^{10}$Department of Physics and Electrical Engineering, Linnaeus University,  351 95 V\"axj\"o, Sweden\\ 
$^{11}$Institut f\"ur Theoretische Physik, Lehrstuhl IV: Weltraum und Astrophysik, Ruhr-Universit\"at Bochum, D 44780 Bochum, Germany\\ 
$^{12}$Institut f\"ur Astro- und Teilchenphysik, Leopold-Franzens-Universit\"at Innsbruck, A-6020 Innsbruck, Austria\\ 
$^{13}$School of Physical Sciences, University of Adelaide, Adelaide 5005, Australia\\ 
$^{14}$LUTH, Observatoire de Paris, PSL Research University, CNRS, Universit\'e Paris Diderot, 5 Place Jules Janssen, 92190 Meudon, France\\ 
$^{15}$Sorbonne Universit\'e, Universit\'e Paris Diderot, Sorbonne Paris Cit\'e, CNRS/IN2P3, Laboratoire de Physique Nucl\'eaire et de Hautes Energies, LPNHE, 4 Place Jussieu, F-75252 Paris, France\\ 
$^{16}$Laboratoire Univers et Particules de Montpellier, Universit\'e Montpellier, CNRS/IN2P3,  CC 72, Place Eug\`ene Bataillon, F-34095 Montpellier Cedex 5, France\\ 
$^{17}$IRFU, CEA, Universit\'e Paris-Saclay, F-91191 Gif-sur-Yvette, France\\ 
$^{18}$Astronomical Observatory, The University of Warsaw, Al. Ujazdowskie 4, 00-478 Warsaw, Poland\\ 
$^{19}$Aix Marseille Universit\'e, CNRS/IN2P3, CPPM, Marseille, France\\ 
$^{20}$Instytut Fizyki J\c{a}drowej PAN, ul. Radzikowskiego 152, 31-342 Krak{\'o}w, Poland\\ 
$^{21}$Funded by EU FP7 Marie Curie, grant agreement No. PIEF-GA-2012-332350\\
$^{22}$School of Physics, University of the Witwatersrand, 1 Jan Smuts Avenue, Braamfontein, Johannesburg, 2050 South Africa\\ 
$^{23}$Laboratoire d'Annecy de Physique des Particules, Univ. Grenoble Alpes, Univ. Savoie Mont Blanc, CNRS, LAPP, 74000 Annecy, France\\
$^{24}$Landessternwarte, Universit\"at Heidelberg, K\"onigstuhl, D 69117 Heidelberg, Germany\\
$^{25}$Universit\'e Bordeaux, CNRS/IN2P3, Centre d'\'Etudes Nucl\'eaires de Bordeaux Gradignan, 33175 Gradignan, France\\
$^{26}$Oskar Klein Centre, Department of Physics, Stockholm University, Albanova University Center, SE-10691 Stockholm, Sweden \\
$^{27}$Institut f\"ur Astronomie und Astrophysik, Universit\"at T\"ubingen, Sand 1, D 72076 T\"ubingen, Germany\\
$^{28}$Laboratoire Leprince-Ringuet, Ecole Polytechnique, CNRS/IN2P3, F-91128 Palaiseau, France\\ 
$^{29}$APC, AstroParticule et Cosmologie, Universit\'{e} Paris Diderot, CNRS/IN2P3, CEA/Irfu, Observatoire de Paris, Sorbonne Paris Cit\'{e}, 10, rue Alice Domon et L\'{e}onie Duquet, 75205 Paris Cedex 13, France\\
$^{30}$Univ. Grenoble Alpes, CNRS, IPAG, F-38000 Grenoble, France\\
$^{31}$Department of Physics and Astronomy, The University of Leicester, University Road, Leicester, LE1 7RH, United Kingdom\\
$^{32}$Nicolaus Copernicus Astronomical Center, Polish Academy of Sciences, ul. Bartycka 18, 00-716 Warsaw, Poland\\
$^{33}$Institut f\"ur Physik und Astronomie, Universit\"at Potsdam,  Karl-Liebknecht-Strasse 24/25, D 14476 Potsdam, Germany\\
$^{34}$Friedrich-Alexander-Universit\"at Erlangen-N\"urnberg, Erlangen Centre for Astroparticle Physics, Erwin-Rommel-Str. 1, D 91058 Erlangen, Germany\\
$^{35}$DESY, D-15738 Zeuthen, Germany\\
$^{36}$Obserwatorium Astronomiczne, Uniwersytet Jagiello{\'n}ski, ul. Orla 171, 30-244 Krak{\'o}w, Poland\\ 
$^{37}$Centre for Astronomy, Faculty of Physics, Astronomy and Informatics, Nicolaus Copernicus University,  Grudziadzka 5, 87-100 Torun, Poland\\
$^{38}$Department of Physics, University of the Free State,  PO Box 339, Bloemfontein 9300, South Africa\\ 
$^{39}$Department of Physics, Rikkyo University, 3-34-1 Nishi-Ikebukuro, Toshima-ku, Tokyo 171-8501, Japan\\
$^{40}$Kavli Institute for the Physics and Mathematics of the Universe (Kavli IPMU), The University of Tokyo Institutes for Advanced Study (UTIAS), The University of Tokyo, 5-1-5 Kashiwa-no-Ha, Kashiwa City, Chiba, 277-8583, Japan \\
$^{41}$Department of Physics, The University of Tokyo, 7-3-1 Hongo, Bunkyo-ku, Tokyo 113-0033, Japan \\
$^{42}$RIKEN, 2-1 Hirosawa, Wako, Saitama 351-0198, Japan \\
$^{43}$Now at Physik Institut, Universit\"at Z\"urich, Winterthurerstrasse 190, CH-8057 Z\"urich, Switzerland \\
$^{44}$Now at Institut de Ci\`{e}ncies del Cosmos (ICC UB), Universitat de Barcelona (IEEC-UB), Mart\'{i} Franqu\`es 1, E08028 Barcelona, Spain\\
$^{\star}$ Corresponding author \\
$^{\dagger}$ Deceased
}

\clearpage

%\maketitle

%% Redefine numeric symbols for footnotes
%\makeatletter
%\renewcommand*{\@fnsymbol}[1]{\ifcase#1\@arabic{#1}\fi}
%\makeatother 

  \begin{abstract}
 PKS 1830-211 is a known macrolensed quasar located at a redshift of z=2.5. Its high-energy gamma-ray emission has been detected with the {\it Fermi-LAT} instrument and evidence for lensing was obtained by several authors from its high-energy data.
   Observations of PKS 1830-211 were taken with the H.E.S.S. array of Imaging Atmospheric Cherenkov Telescopes in August 2014, following a flare alert by the {\it Fermi- LAT} collaboration. The H.E.S.S observations were aimed at detecting a gamma-ray flare delayed by 20-27 days from the alert flare, as expected from  observations at other wavelengths.
   More than twelve hours of good quality data were taken with an analysis threshold of $\sim67$ GeV. The significance of a potential signal is computed as a function of the date as well as the average significance over the whole period. Data are compared to simultaneous observations by {\it Fermi-LAT}.
   No photon excess or significant signal is detected. An upper limit on PKS 1830-211 flux above 67 GeV is computed and compared to the extrapolation of the {\em Fermi-LAT} flare spectrum. 
\end{abstract}
  
%     \keywords{gamma-ray astronomy --
 %                      gravitational lensing--
 %                      PKS 1830-211

\section{The PKS 1830-211 gravitationally lensed quasar}\label{sec:introduction}
PKS 1830-211 is a high redshift (z=2.5 \citep{1999ApJ...514L..57L}) Flat Spectrum Radio Quasar (FSRQ) which has been detected in all wavelengths from radio to high-energy gamma-rays. It is a known gravitationally lensed  object with two compact images of the quasar nucleus visible in the radio  \citep{1991Natur.352..132J} and optical \citep{2005A&A...438L..37M} passbands. The Einstein ring, well visible
at radio frequencies, comes from the imaging of the quasar jet \citep{1992ApJ...401..461K}.   
The quasar source is lensed by a foreground galaxy at  z=0.89 \citep{1996Natur.379..139W}. The angular size of the 
Einstein ring and separation of compact images is roughly 1 arcsecond so that it cannot be resolved with high-energy instruments such as H.E.S.S. (50 GeV-50 TeV range) or {\it Fermi-LAT} (100 MeV-100 GeV range). PKS 1830-211 is seen as a bright, high-energy source by the {\em Fermi-LAT} instrument and had several flaring periods during the decade of {\em Fermi-LAT} observations.  
%PKS 1830-211 is listed in the 1FHL catalogue \citep{2013ApJS..209...34A} with a spectral index above 10 GeV of $\sim 3.9,$ which corresponds to the average "low-state" spectrum. 
PKS 1830-211 is listed in the 1FHL \citep{2013ApJS..209...34A} and the 3FHL \citep{2017ApJS..232...18A}  catalogues with a 
%spectral
photon index above 10 GeV of $3.55 \pm  0.34$ which corresponds to the average "low-state" spectrum. 
No significant curvature in the spectrum was detected. Photons up to 35 GeV, potentially detectable by H.E.S.S. have been observed by {\em Fermi-LAT} \citep{2017ApJS..232...18A}. 
Observations of these very high energy photons and the measurement of the very high energy tail of the spectrum would give useful constraints on EBL at redshift z=2.5.

Since the components of the lens cannot be resolved at high or very high energy, the evidence for lensing was searched indirectly on the observed light curve. Due to the different travel paths, the 
light curves of the two compact components of the lens have a relative time delay, measured in the radio \citep{1998ApJ...508L..51L} and  microwave \citep{2001ASPC..237..155W} pass-bands, of  $26\pm5$ days. 
\cite{2011A&A...528L...3B} have studied the first three years of the {\em Fermi-LAT} light curve with cepstral and autocorrelation methods. 
Evidence for a delay of  $27.5\pm1.3$ days was found with a 3 $\sigma$ significance. The time delay between the compact images of PKS 1830-211 was also studied by the {\em Fermi-LAT} collaboration \citep{2015ApJ...799..143A}. They selected  several flaring periods and calculated the autocorrelation function of the light curve. No significant peak was found. A possible peak of  
$\sim 20$ days was found with a 1-day binning of the data, which could be attributed to the $\sim 20$ days separation between two flaring events and perhaps to gravitational lensing. 
\citet{2015ApJ...809..100B} have argued that the time delay measured by high-energy instruments could be very different than the value measured by radio telescopes. The delay measured by 
\cite{1998ApJ...508L..51L} is obtained from the emission of the compact images. 
Since the jet of the PKS 1830-211 source is imaged close to the Einstein ring, the time difference
between the intial burst and its lensed image can be much smaller if the source of high-energy emission is located inside the jet.

% Parler de Neronov et al. 
%The {\em Fermi-LAT} collaboration has claimed the discovery of a gravitationally induced delay on another  known gravitational lens object, the B0218+357  FSRQ  \citep{2014ApJ...782L..14C}. The high energy time delay ($\sim 11$ days) was measured to be slightly longer than, but compatible with the radio time delay. The lensed AGN, located at z=0.94, was also detected and studied with the MAGIC Imaging Atmospheric Cherenkov Telescopes (IACTs) \citep{2014ATel.6349....1M}.  MAGIC was able to detect a delayed flare 11 days after an alert sent by the {\em Fermi-LAT} collaboration. The delayed flare was neither observed by {\em Fermi-LAT}, nor in optical or X-ray passbands \citep{2016A&A...595A..98A}.  
PKS 1830-211 is monitored by {\em Fermi-LAT} and its light curve is posted on the internet
\footnote{\tiny https://fermi.gsfc.nasa.gov/ssc/data/access/lat/msl\_lc/source/PKS\_1830-211}
on a daily basis. 
H.E.S.S. observations of PKS 1830-211 were triggered by an alert posted by the {\em Fermi-LAT} team on August 2, 2014 \citep{2014ATel.6361....1K}. The flare seen by the {\em Fermi-LAT} instrument started on July 27 and lasted $\sim$4 days.  The H.E.S.S. observations are described in Section \ref{sec:observations} and data analysis in Section \ref{sec:analyses}. The H.E.S.S. limits are compared to the 
{\em Fermi-LAT} signal in Section \ref{sec:limits} and discussed in Section \ref{sec:conclusion}. 
\begin{figure}
\centering
\includegraphics[height=6cm]{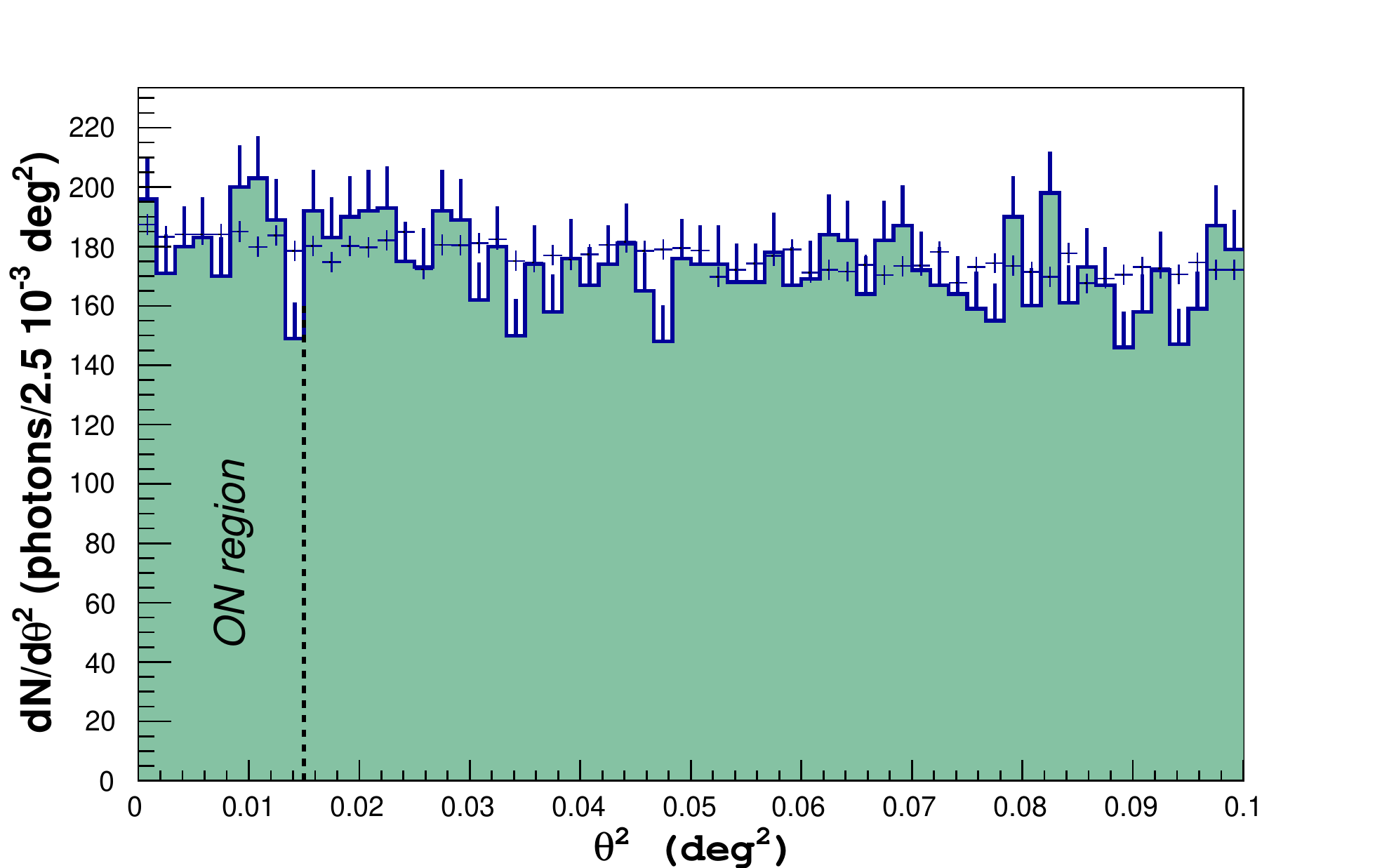}
\caption{$\theta^2$ plot of PKS 1830-211 obtained with the {\em Mono} reconstruction. The background, shown by crosses, is estimated with the ring background method.
% and the significance of the excess was computed with the formula of \citet{1983ApJ...272..317L}. 
%Right: H.E.S.S. significance map of PKS 1830-211. The background is substracted by the  ring background method. 
}
\label{fig:hess-obs1}
\end{figure}

\section{H.E.S.S. observations}\label{sec:observations}
The very-high-energy (50 GeV-50 TeV range) gamma-ray observatory of the H.E.S.S. collaboration consists of
five  Imaging Atmospheric Cherenkov Telescopes (IACTs) located in the Khomas Highland 
of Namibia ($23 ^{\circ}$ 16' 18'' S, $16 ^{\circ}$ 30' 1'' E), 1800 m above sea level. 
From January 2004 to October 2012, the array was a four-telescope instrument, with telescopes labeled CT1-4. 
 Each of the telescopes, located at the corners of a square with a side length of 120 m has an effective mirror 
surface area of 107~m$^{2}$, and is able to detect cosmic gamma-rays in the energy 
range 0.1 -- 50 TeV .
In October 2012, a fifth telescope CT5, with an effective mirror surface area of 600~m$^{2}$ and an improved camera \citep{2014NIMPA.761...46B} was installed at the center of the original square, giving access to energies below 100 GeV \citep{2017A&A...600A..89H}. 

PKS 1830-211 was observed by the five telescopes of the H.E.S.S. IACT array between August 12 2014 and August 26 2014, to allow for the detection of delayed flares with time delays ranging from 20 to 27 days. 
The observations were taken at an average zenith angle of 12 degrees. 
%The data were taken with the five telescopes.
%but the trigger rate was dominated by CT5.

\section{Data analyses}\label{sec:analyses}
 This paper is based on a sample of 12.4 hours of high quality data. Data selection cuts have been described in
 \citep{2017A&A...600A..89H}.
  Data were next analyzed with the Model analysis \citep{2009APh....32..231D} and cross-checked with the ImPACT analysis \citep{2014APh....56...26P}, the two methods giving compatible results.  The two analyses use different calibration chains. 
 With both reconstruction chains, data of CT5 were analyzed either alone ({\em Mono} reconstruction) or combined with the CT1-4 data ({\em Combined} reconstruction).   The {\em Mono} reconstruction has an energy threshold of 67 GeV. 
% Please check.
The {\em Combined} reconstruction has a higher threshold of 144 GeV,
% Please check
 but a larger effective area. 

A point source is searched at the location of PKS 1830-211.  
Fig. \ref{fig:hess-obs1} shows the distribution of the squared angular distance $\theta^2$ of candidate photons from the target position. This distribution, obtained in the {\em Mono} analysis, is compared to 
the background from hadrons mis-identified as photons. The background is calculated with the {\em ring background} method \citep{2007A&A...466.1219B}, other methods giving similar results. 
%$\theta$ is the angular distance from the target position.
% The background from hadrons mis-identified as photons was calculated with several methods, giving identical results. 

%No significant point source is found at the PKS 1830-211 location with the {\em Combined} analysis. 
Table \ref{tab:results} summarizes the number of candidate photons in the signal region, the expected background and the significance of the excess,  calculated with Li and Ma formula 17 \citep{1983ApJ...272..317L}. 
\begin{table}
\caption{Analysis results of observations of PKS 1830-211 by H.E.S.S.}
\label{tab:results}
\centering
\begin{tabular}{c c c c}
\hline \hline
Reconstruction& $N_\mathrm{ON}$ & $N_\mathrm{background}$ & significance ($\sigma$) \\
\hline
{\em Mono} & 1641 & 1649.2 & -0.2 \\
{\em Combined} & 935 & 954.4 & -0.6 \\
\hline
\end{tabular}
\end{table}
%A signal of 686 candidate photons was observed, while the background estimated by the ring background method is 700.2. 
% The significance of the photon excess is -0.5 $\sigma$ and was calculated with Li and Ma formula 17 \citep{1983ApJ...272..317L}. 
No significant excess of photons over background is seen by H.E.S.S. at the position of PKS 1830-211.  A similar search using the {\em Combined} analysis also gives a negative result.

\begin{figure}
\centering
\includegraphics[height=4.5cm]{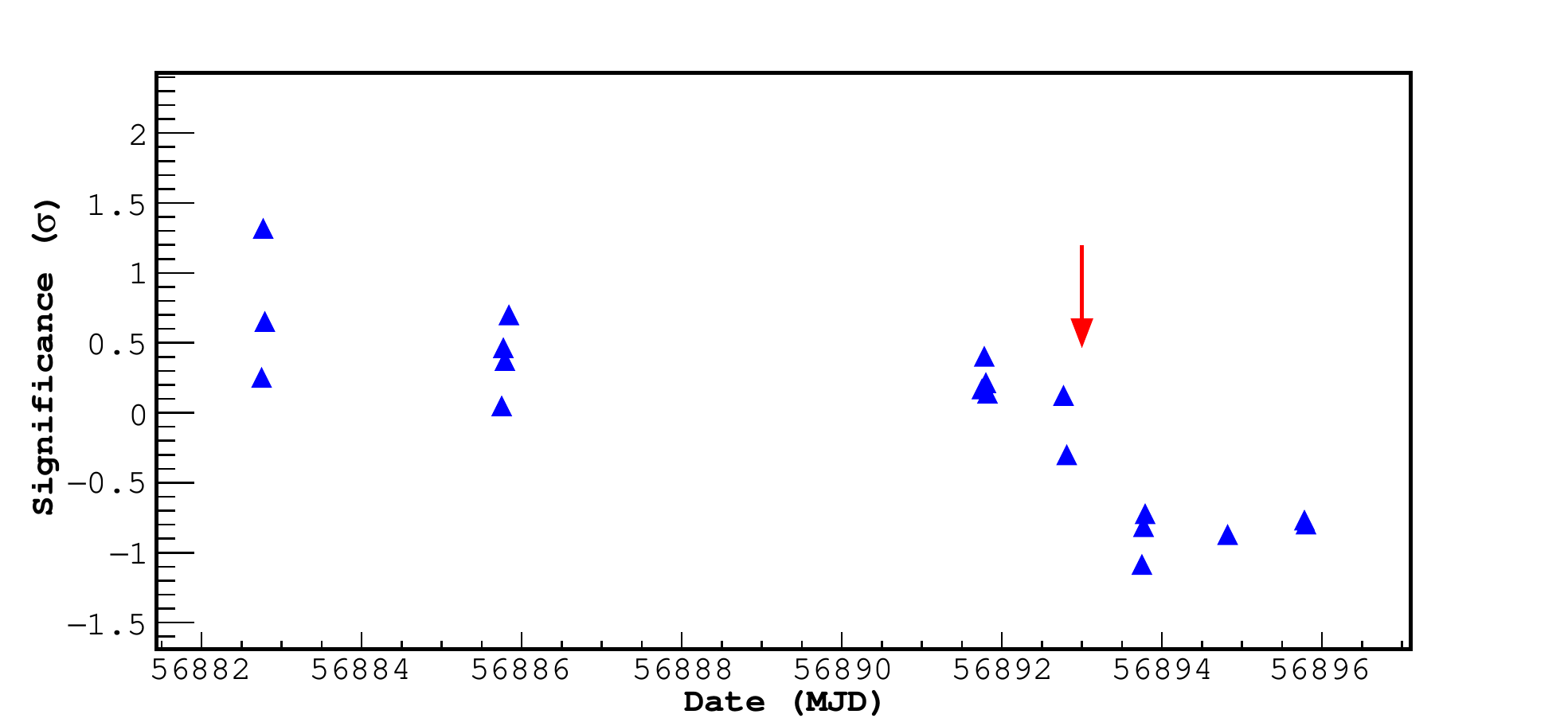}
\caption{
%Left: Excess signal in the PKS 1830-211 region vs. livetime
%for H.E.S.S. August 2014 observations. Right: 
Significance of the H.E.S.S. signal versus date, obtained with the {\em Mono} analysis. The red arrow shows the expected date of the delayed flare for a lensing time delay 
of 27 days. 
%H.E.S.S. data are binned by 28 minute runs.
%The significance is computed according to the recipe of Li and Ma. 
%The delayed
%Fermi flare (red arrow) would be expected around MJD 56893 for the time delay between flares of 27 days.
}
\label{fig:hess-obs2b}
\end{figure}

Because of the very soft spectrum measured by {\em Fermi LAT} in the low state, PKS~1830-211 has a chance of being detectable by H.E.S.S. only during flares. 
The delayed flare lasts only less than about 4 days, however, due to the uncertainties on the date of the flare, it could have happened at any time 
between August 17 = MJD 56886  (time delay of 20 days) and August 24  = MJD 56893 (radio time delay of 27 days) as explained in Section \ref{sec:introduction}. 
Fig. \ref{fig:hess-obs2b} shows the evolution over time of significance, binned by 28-minute runs.
No significant daily photon excess was detected during the H.E.S.S. observation period. 

%\begin{figure}[h]
%\centering
%\includegraphics[height=6cm]{sig-vs-time-icrc-v4-prel.png}
%\caption{
%Excess signal in the PKS 1830-211 region vs. livetime
%for H.E.S.S. August 2014 observations. 
%Right: Significance of the H.E.S.S. signal versus date. The significance is computed according to the recipe of Li and Ma. The delayed
%Fermi flare would be expected around MJD 56886 if the time delay between flares was 20 days.
%}
%\label{fig:hess-obs2a}
%\end{figure} 

\section{Flux upper limits and comparison to the {\em Fermi-LAT} spectra}\label{sec:limits}
\begin{figure}
\centering
\includegraphics[width=9cm]{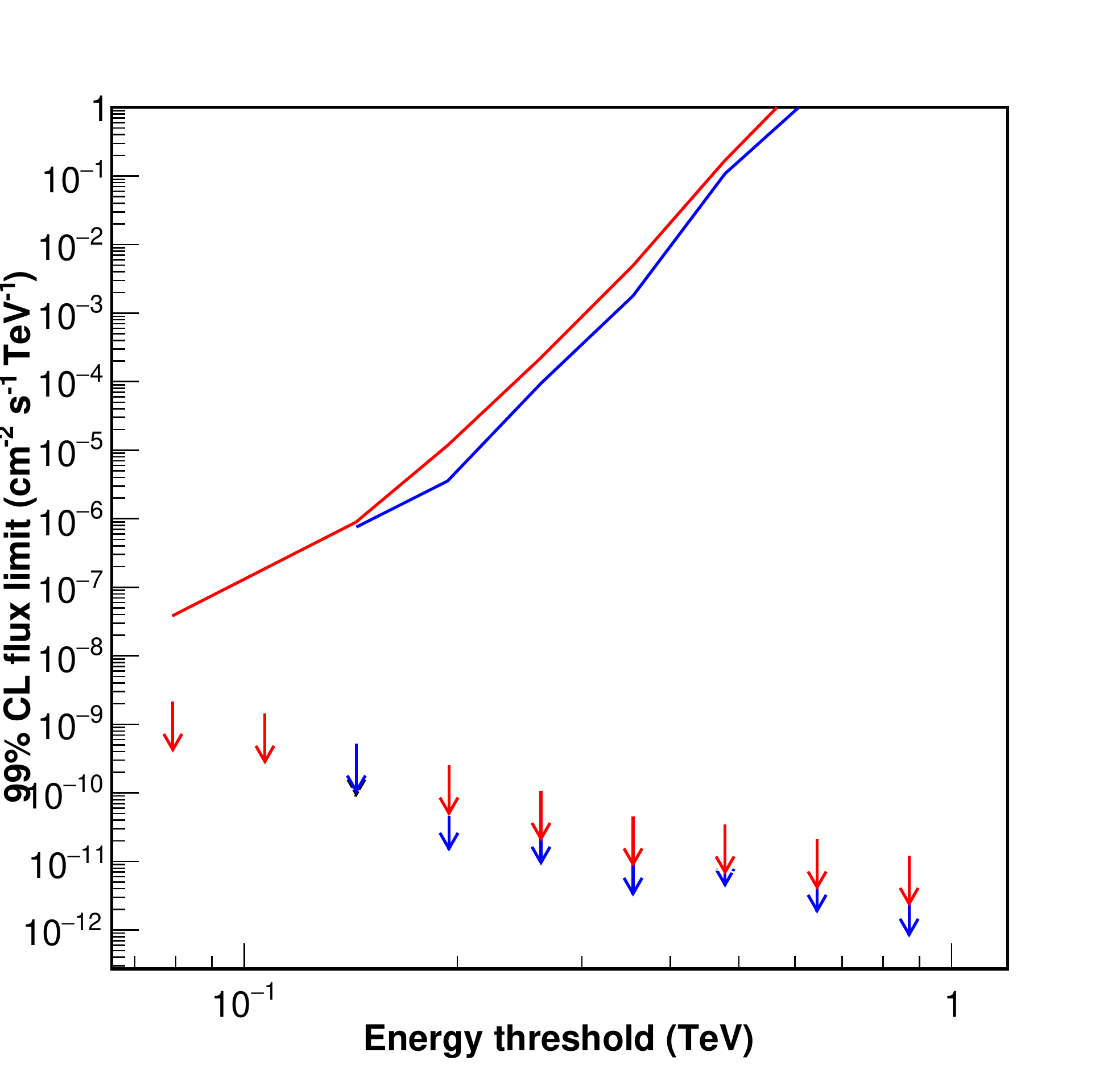}
\caption{99\% C.L. upper limits  (arrows) on the PKS 1830-211 flux between 67 GeV and 1 TeV for the August 2014 H.E.S.S. observations.  A constant photon index of -3 was assumed. The solid lines show the effect of EBL deabsorption, assuming the EBL model of \citet{2012MNRAS.422.3189G}.  
}
\label{fig:hess-ul}
\end{figure}

%The null result of the H.E.S.S. observations 
The non-detection by H.E.S.S. 
translates into 99\% confidence level (C.L.) upper limits  on the average very-high-energy flux of PKS 1830-211 during H.E.S.S. observations. These upper limits are shown in Fig. \ref{fig:hess-ul}. Red (resp. blue) arrows 
show the limits obtained from the {\em Mono} (resp. {\em Combined}) analysis and the corresponding solid lines show the effect of deabsorption using the Extragalactic Background Light (EBL) model of  \citet{2012MNRAS.422.3189G}. 

\begin{figure}
\centering
\includegraphics[width=9cm]{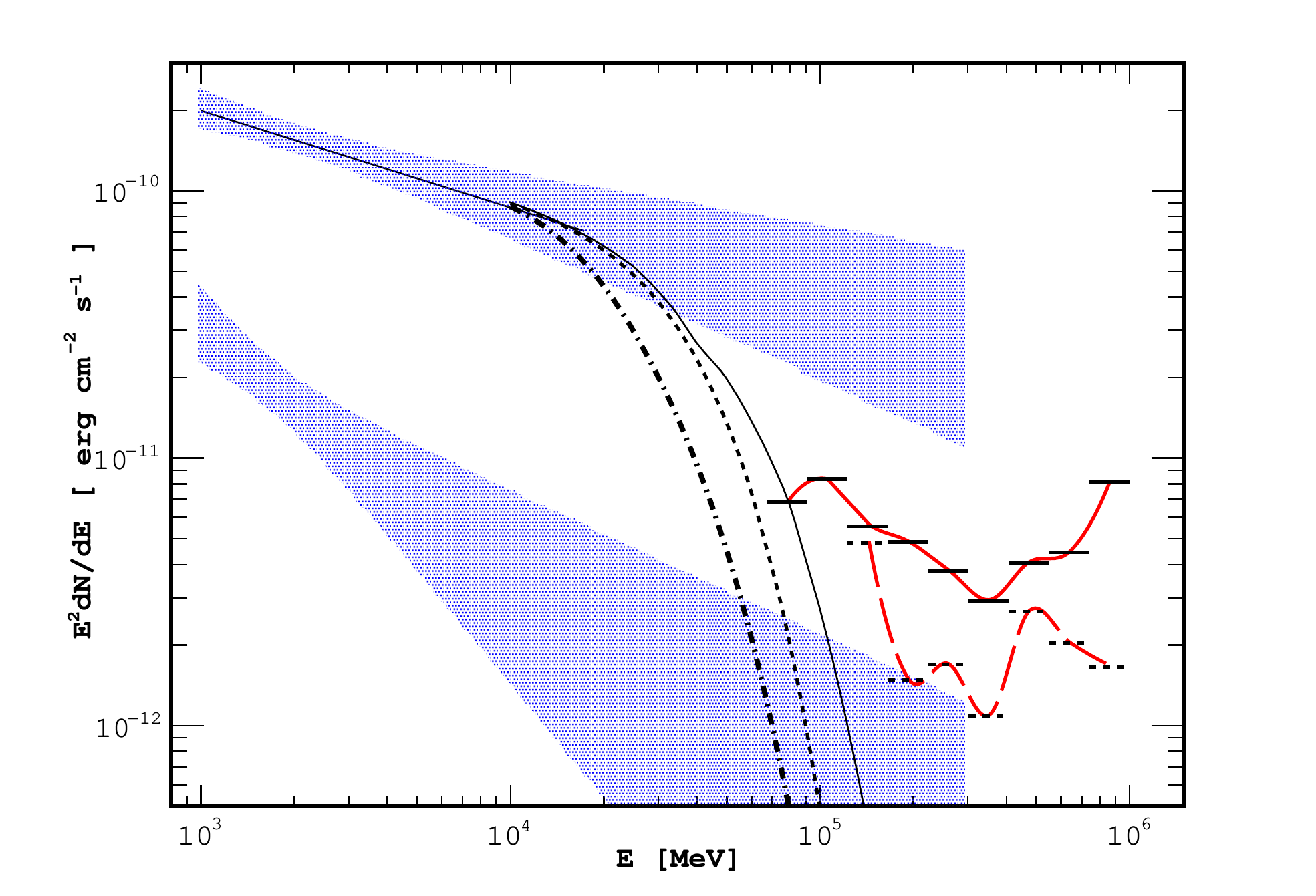}
\caption{Comparison of H.E.S.S. flux 99 \%C.L. upper limits (red solid line: {\em Mono} analysis, long dashed line: {\em Combined} analysis) to the measured
spectra in the GeV region obtained with the {\em Fermi-LAT} data. The horizontal lines show the spectral resolution of the analyses. The lower blue butterfly is the GeV spectrum of PKS 1830-211 during H.E.S.S. observations. The upper blue butterfly is the 
corresponding spectrum during the July 2014 flare.  The absorption of the July flare spectrum by EBL is calculated with models by \citet{2012MNRAS.422.3189G} (black dash-dotted line),\citet{2010ApJ...712..238F} (dotted line) and \citet{2008A&A...487..837F} (black solid line).  
%The solid line gives the expected limit from H.E.S.S. observations.
}
\label{fig:hess-fermi-comp}
\end{figure}

H.E.S.S. upper limits are compared to {\em Fermi-LAT} GeV spectra on Fig \ref{fig:hess-fermi-comp}. 
The {\em Fermi-LAT} observations have been analyzed with Fermi Science Tools v10-r0p5
and Pass8 data, in the Enrico framework \citep{2013arXiv1307.4534S}. The spectral data from 26-30 July 2014 (flare) are well 
described by a power-law spectrum with an index of $n_{flare}= -2.36 \pm 0.17$ for photon energies $> 1$ GeV. The relatively high low-energy cut was used to avoid contamination from the Galactic plane. 
 The flare spectrum is much harder than the spectrum measured in the low state of PKS 1830-211, but $n_{flare}$ is compatible with the photon indices of previous flare spectra, as measured by {\em Fermi-LAT}  \citep{2015ApJ...799..143A}. The 
spectrum of PKS 1830-211 obtained from the {\em Fermi-LAT} observations within the H.E.S.S. observation window is well described by a power-law with an index of $n_{low}=-2.97 \pm 0.44$ above 1 GeV.  The value of $n_{low}$ is compatible with the value published in the 3FHL catalogue.

%H.E.S.S. energy flux upper limits are above the extrapolation of the low-state spectrum measured by {\em Fermi-LAT} and below the extrapolation of the flare spectrum.  However, a 
A proper comparison between H.E.S.S. upper limits and the Fermi signal has to take into account the effect of the absorption of the flux of PKS 1830-211 by the 
%extragalactic background light
 EBL and the difference between the flare duration and H.E.S.S. exposure. Since no significant curvature of the spectrum was measured bt the Fermi-LAT collaboration, the unabsorbed spectrum was modeled by a powerlaw.
  The effect of light absorption by EBL from PKS 1830-211 has been estimated with the models of  
 \citet{2012MNRAS.422.3189G} (black dash-dotted line), \citet{2010ApJ...712..238F} (dotted line) and \citet{2008A&A...487..837F} (black solid line).  Fig. \ref{fig:hess-fermi-comp} shows that there 
 is a substantial difference between the predictions of these models for a source at redshift 2.5 such as PKS 1830-211. 
 Note that EBL absorption could also be affected by the lens environment.
 Light from lensed AGN is expected to be more absorbed than average, due to the presence of galaxies along the line of sight. Indeed, absorption from the intervening galaxy has been detected by \cite{2005A&A...438..121D} in the X-ray spectrum of PKS 1830-211. However \cite{2014ApJ...790..147B} and \cite{2016A&A...595A..14B}  have argued that gravitational lensing 
 could  help gamma-rays from a distant source avoiding excess absorption. 
 The {\em Fermi-LAT} flare spectrum from Fig \ref{fig:hess-fermi-comp} is a 4-night average while the H.E.S.S. exposure amounts to 10 nights of data taking. The steady source upper limits from Fig \ref{fig:hess-ul} are thus  a factor $\sim\sqrt{10/4}$ too constraining, which is corrected for in Fig   \ref{fig:hess-fermi-comp}. 
 
% Fig. \ref{fig:hess-fermi-comp}  shows the effect of EBL calculated with 2 different models, from \citet{2012MNRAS.422.3189G} and \citet{2010ApJ...712..238F}. The energy threshold of H.E.S.S. data is too high to detect  the PKS 1830-211 bursts with standard EBL models.  
 
\section{Conclusion}\label{sec:conclusion} 
 
No significant delayed flare from PKS 1830-211 was detected by either H.E.S.S. or {\em Fermi-LAT}. The flare did not repeat or was too faint to be detected.
Fig \ref{fig:hess-fermi-comp} shows however that the detection of a strong flare would have been possible 
%if EBL absorption is accurately described 
close to the {\em Mono} analysis energy threshold if the level of EBL absorption was at or below the absorption predicted
by the model of  \citet{2008A&A...487..837F}.
Due to its lensed nature, observation of flaring event of PKS 1830-211 in the TeV passband could be useful to constrain EBL models at redshift as large as  2.5. 
%In the future, H.E.S.S. observations during a {\em Fermi-LAT} alert could be useful to constrain  EBL models at a redshift of 2.5.
%since  {\em Fermi-LAT} observations do not strongly constrain models beyond a redshift of 2 \citep{2017sf2a.conf..155B}. 
%If in addition the echoed flare is detected,  H.E.S.S. could accurately measure 
The detection of the lensing time delay in future very high energy observations
%If there had been an echoed flare, H.E.S.S. would have in any case been (very) useful. It could have either accurately measure the lensing time delay at very high energy in case of a detection or put constraints 
%on EBL models at a redshift of 2.5 in case of a non-detection. 
%In case of a non-detection by H.E.S.S. of an echoed flare detected by {\em Fermi-LAT}, H.E.S.S. would have put significant constraints on EBL models at a redshift of 2.5.   
 %In case of a detection, H.E.S.S. could accurately measure the lensing time delay at very high energy.
 would help pinpoint the spatial origin of the high-energy emission \citep{2015ApJ...809..100B}.
 It would also permit more exotic applications such as constraining photon mass \citep{2017ApJ...850..102G} or testing Lorentz Invariance Violation \citep{2009MNRAS.396..946B}.

\section*{Acknowledgments}
{\small
The support of the Namibian authorities and of the University of Namibia in facilitating the construction
 and operation of  H.E.S.S. is    gratefully acknowledged, as is the support by the German Ministry for Education and Research (BMBF), 
 the Max Planck Society, the German Research Foundation (DFG), the Helmholtz Association, the Alexander von Humboldt Foundation,
the French Ministry of Higher Education, Research and Innovation, the Centre National de la Recherche Scientifique (CNRS/IN2P3 and CNRS/INSU), 
the Commissariat  \`a    l' Energie Atomique et aux Energies Alternatives (CEA), 
the  U.K. Science and Technology Facilities Council (STFC), the Knut and Alice Wallenberg Foundation, the National Science Centre, 
 Poland grant no. 2016/22/M/ST9/00382, the South African Department of Science and Technology and National 
Research Foundation, the University of Namibia, the National Commission on Research, Science \&  Technology of  Namibia (NCRST), the Austrian Federal Ministry of Education, Science and Research 
and the Austrian Science Fund (FWF), the Australian Research Council (ARC), the Japan Society for the Promotion of Science and by the University of Amsterdam. 
We appreciate the excellent work of the technical
 support staff in Berlin, Zeuthen, Heidelberg, Palaiseau, Paris, Saclay, T\"uebingen and in Namibia in the construction and operation of the equipment. This work benefited from services provided by the
H.E.S.S. Virtual Organisation, supported by the national resource providers of the EGI Federation
}

%\begin{thebibliography}{99}
%\bibitem{ackermann} M. Ackermann et al. (Fermi-Lat collaboration), ApJS, 209, 34 (2013)
%\bibitem{lovell}J.E.Lovell, D.L.Jauncey, J.E.Reynolds et al., ApJ. 58, L51 (1998)
%\bibitem{combes}T.Wiklind, F.Combes, Proceedings ASPC, 237, 155 (2001)
%\bibitem{barnacka1} A.Barnacka, J-F. Glicenstein, Y. Moudden A\&A Letters, 728, 3 (2011) 
%\bibitem{abdo} A.A Abdo et al. (Fermi-LAT collaboration), ApJ 799, 143 (2015)
%\bibitem{cheung} C.C. Cheung,S. Larsson, J.D. Scargle, et al., ApJ, 782, L14 (2014)
%\bibitem{mirzoyan}R. Mirzoyan, The Astronomer's Telegram, 6349, 1 (2014)
%\bibitem{lima} T. Li, Y. Ma, ApJ. 272, 317 (1983)
%\bibitem{krauss}F. Krauss, J.Becerra, B. Carpenter et al., The Astronomer's Telegram,  6361, 1 (2014)
%\bibitem{sanchez}D.A.Sanchez, C. Deil, arXiv:1307.4534 (2013)
%\bibitem{barnacka2}A. Barnacka, M.B\"ottcher, I. Sushch, ApJ. 790, 147 (2014)
%\bibitem{holler}M. Holler, A. Balzer, M. de Naurois et al., these proceedings, ICRC2015-I/509 (2015)
%\bibitem{parsons}R. Parsons, T. Murach, M. Gajdus, these proceedings, ICRC2015-I/559 (2015) 
%\end{thebibliography}
\bibpunct{(}{)}{;}{a}{}{,} % to follow the A&A style
\bibliographystyle{mnras}
\bibliography{PKS1830}{}

\end{document}